\documentstyle[12pt]{article}
\title{\bf Field Correlators\\
in the Georgi-Glashow Model and\\
Several Aspects
of its String Representation}
\author{D.V.ANTONOV \thanks{E-mail addresses: 
antonov@pha2.physik.hu-berlin.de and antonov@vxitep.itep.ru}{\,}
\thanks{Supported by Graduiertenkolleg {\it Elementarteilchenphysik}, 
Russian Fundamental Research Foundation, Grant No.96-02-19184, DFG-RFFI, 
Grant 436 RUS 113/309/0, and by the INTAS, Grant No.94-2851.}
\\
{\it Institute of Theoretical and Experimental Physics,}\\
{\it B.Cheremushkinskaya 25, 117 218, Moscow, Russia}\\
{\it and}\\
{\it Institut f\"ur Physik, Humboldt-Universit\"at zu Berlin,}\\
{\it Invalidenstrasse 110, D-10115, Berlin, Germany}} 
\date{}
\begin{document}
\maketitle
\vspace{1mm}
\centerline{\bf {Abstract}}
\vspace{3mm}
String representations of the Wilson loop and of the non-Abelian analog 
of the 't Hooft loop, defined on the string world-sheet, are studied 
in the Londons' limit of the 4D Georgi-Glashow model. 
In the first case, 
massive gauge bosons yield only perimeter type terms, whereas in the 
second case, they also interact with the string, which 
enables one to derive the coefficient functions  
parametrizing the bilocal correlator of the dual field strength 
tensors. The asymptotic behaviours of these functions 
at small and large distances are 
argued to be in a good agreement with the observable lattice data.

\vspace{6mm} 

In the recent papers$^{1,2}$, the string representation for the 't Hooft 
loop average defined on the string world-sheet 
in the Londons' limit of the Abelian Higgs Model (AHM) has been 
found. In particular in Ref. 1, it has been demonstrated that the 
bilocal approximation to the Method of Vacuum Correlators (MVC)$^{3}$ 
is exact 
in the Londons' limit, i.e. all the terms of the cumulant expansion 
higher than the quadratic one vanish. The value of the bilocal 
correlator has been 
derived from the propagator of the massive Kalb-Ramond field, and 
is a good agreement 
with the present lattice data concerning this correlator in 
gluodynamics$^{4}$.

In this Letter, we shall proceed to the non-Abelian case of the 
4D Georgi-Glashow model and study string representations for the 
Wilson loop average and the non-Abelian analog of the 't Hooft loop
average defined on the string world-sheet. 
To this end, we shall make use of the same 
duality transformation, which has been originally proposed in Ref. 5 and 
then used in Refs. 1 and 2, and earlier in 6 and 7. 
In this way, we shall finally 
also find the bilocal correlator of the dual non-Abelian field strength 
tensors. All the calculations will be performed in the 
Londons' limit of the Georgi-Glashow 
model. Notice however, that contrary to Ref. 7, where only the low-energy 
limit of the Georgi-Glashow model (which is in fact an Abelian theory) 
has been considered, in what folllows we shall study 
this model as it is, i.e. the non-Abelian case.

We shall start with the string representation of the  
Wilson loop average 

$$\left<W(C)\right>=\left<{\rm tr}{\,}{\rm P}{\,}\exp\left(iq\oint
\limits_C^{}
dx_\mu A_{\mu}^aT^a\right)\right>,$$
formed by a probe external electric charge $q$ moving along a closed 
contour $C:{\,}x_\mu(\tau),{\,}0\le\tau\le 1$, and belonging to the 
representation of the group $SU(2)$ with the colour ``spin'' $J$, i.e. 
$\left(T^a\right)^{bc}\left(T^a\right)^{cd}=J(J+1)\delta^{bd}$ 
(for example, $J=1$ for the adjoint representation). 
To this end, we shall make use of the formula derived in Ref. 8, which 
enables one to replace the path ordering in the Wilson loop in the 
$SU(2)$ gauge theory by the integration over an auxiliary unit 
three-vector. Note, that this formula has already been applied for 
the purposes 
of the string representation of gluodynamics in the confining background 
in Ref. 9. So, the Wilson loop average we are starting with 
reads as follows

$$\left<W(C)\right>=\frac{1}{{\cal Z}}
\int {\cal D}A_\mu^a {\cal D}\Phi^a\Biggl<
\exp\Biggl[\int dx\Biggl(-\frac14 F_{\mu\nu}^{a2}-\frac12\left|
D_\mu^{ab}\Phi^b\right|^2-\lambda\left(\left|\Phi^a\right|^2-\eta^2
\right)^2+iJqj_\mu^aA_\mu^a\Biggr)\Biggr]\Biggr>_{\vec n}, \eqno (1)$$
where

$$\left<...\right>_{\vec n}\equiv\int {\cal D}\vec n \left(...\right)
{\rm e}^{iJ\int\limits_0^1 d\tau\dot\phi\cos\chi}$$
with the vector $\vec n$ being parametrized as 

$$\vec n=\left(\sin\chi\cos\phi,{\,}\sin\chi\sin\phi,{\,}\cos\chi\right),
~{\rm and}~ j_\mu^a(x)=\int\limits_0^1d\tau\dot x_\mu(\tau)n^a\left(x(\tau)
\right)\delta(x-x(\tau))$$ 
being its current along the contour $C$. 
The covariant derivative and the non-Abelian field 
strength tensor are defined as 
$D_\mu^{ab}=\delta^{ab}\partial_\mu-g\varepsilon^{abc}A_\mu^c$ 
and $F_{\mu\nu}^a=\partial_\mu A_\nu^a-\partial_\nu A_\mu^a+g
\varepsilon^{abc}A_\mu^bA_\nu^c$ respectively.

In the Londons' limit, when the coupling 
constant $\lambda=+\infty,{\,}\Phi^a(x)=N^a\eta {\rm e}^{i\theta(x)}$,  
where $\theta=\theta^{{\rm sing.}}+\theta^{{\rm reg.}}$, and 
$\vec N$ stands for a {\it constant} unit three-vector, whose arbitrary 
direction must be averaged over. In this limit, Eq. (1) reads

$$\left<W(C)\right>=\frac{1}{{\cal Z}}
\int {\cal D}A_\mu^a {\cal D}\theta\Biggl<
\exp\Biggl[\int dx\Biggl(-\frac14 F_{\mu\nu}^{a2}-\frac{\eta^2}{2}
\left(\partial_\mu\theta\right)^2-\frac{g^2\eta^2}{3}A_\mu^{a2}+iJq 
j_\mu^aA_\mu^a\Biggr)\Biggr]\Biggr>_{\vec n}, \eqno (2)$$
where the coefficient at the mass term of the field 
$A_\mu^a$ is just due to the 
averaging over the directions of the vector $\vec N$. Performing the 
duality transformation proposed in Ref. 5, we get from Eq. (2)

$$\left<W(C)\right>=\frac{1}{{\cal Z}}
\int {\cal D}x_\mu(\xi) {\cal D}h_{\mu\nu}
\exp\Biggl[\int dx\Biggl(-\frac{1}{12\eta^2}H_{\mu\nu\lambda}^2+
i\pi h_{\mu\nu}T_{\mu\nu}\Biggr)\Biggr]\cdot$$

$$\cdot\Biggl<\int {\cal D}A_\mu^a 
\exp\Biggl[\int dx\Biggl(-\frac14 F_{\mu\nu}^{a2}-\frac{g^2\eta^2}{3}
A_\mu^{a2}+iJqj_\mu^aA_\mu^a\Biggr)\Biggr]\Biggr>_{\vec n}, \eqno (3)$$
where $H_{\mu\nu\lambda}=\partial_\mu h_{\nu\lambda}+\partial_\lambda 
h_{\mu\nu}+\partial_\nu h_{\lambda\mu}$ stands for the 
strength tensor of the massless
Abelian antisymmetric tensor field $h_{\mu\nu}$, $T_{\mu\nu}(x)=
\int\limits_S^{} 
d\sigma_{\mu\nu}(x(\xi))\delta(x-x(\xi))$ is the vorticity tensor 
current, $S$ is a closed world-sheet of the string 
parametrized 
by $x_\mu(\xi)$, and from now on, for simplicity, we shall not take into 
account the Jacobian emerging when one passes from the integration over 
$\theta^{{\rm sing.}}$ to the integration over $x_\mu(\xi)$. 
Similarly to Ref. 5, 
the first line in Eq. (3) represents the interaction of a massless photon 
with the string world-sheet. Due to the invariance of this 
expression under the hypergauge transformations, $h_{\mu\nu}\to 
h_{\mu\nu}+\partial_\mu\Lambda_\nu-\partial_\nu\Lambda_\mu$, some 
gauge fixing term is needed for carrying out the integral over the 
field $h_{\mu\nu}$. For example, if one adds to the Lagrangian standing 
in the exponent in the first line of Eq. (3) the mass term $-\frac{1}{4e^2}
h_{\mu\nu}^2$, the integral over the field $h_{\mu\nu}$ yields (up to 
a factor $\pi^2$) Eq. (2) of Ref. 10. In particular, when the regularizing 
photon mass $\frac{\eta}{e}$ is infinitesimally small, one can show 
that the result of the integration over the field $h_{\mu\nu}$ reads as

$$\exp\Biggl[-\frac{e^2}{2}\oint\limits_{\partial S}^{} dx_\mu \oint
\limits_{\partial S}^{}dy_\mu\frac{1}{\left(x-y\right)^2}\Biggr],$$
(i.e. is just the one of the photodynamics case) and thus vanishes, since 
$\partial S=0$. 
Therefore, the 
nonvanishing (i.e. not reducible to the boundary terms) contribution 
to the string effective action could arise only due to the finite 
photon mass, which plays the role of a hypergauge fixing parameter.

However actually, the first line in Eq. (3) drops out after the division 
by ${\cal Z}$, so that the problem of hypergauge fixing does not 
arise at all, and we arrive at the following expression for 
the Wilson loop average 

$$\left<W(C)\right>=\frac{\Biggl<\int {\cal D}A_\mu^a 
\exp\Biggl[\int dx\Biggl(-\frac14 F_{\mu\nu}^{a2}-\frac{g^2\eta^2}{3}
A_\mu^{a2}+iJqj_\mu^aA_\mu^a\Biggr)\Biggr]\Biggr>_{\vec n}}
{\int {\cal D}A_\mu^a\exp\Biggl[-\int dx\Biggl(\frac14 F_{\mu\nu}^{a2}
+\frac{g^2\eta^2}{3}A_\mu^{a2}\Biggr)\Biggr]}.$$
The Gaussian part of this expression reads

$$\frac{\int {\cal D}A_\mu^a\exp\Biggl[\int dx\Biggl(-\frac14
\left(\partial_\mu A_\nu^a-\partial_\nu A_\mu^a\right)^2-
\frac{g^2\eta^2}{3}A_\mu^{a2}+iJqj_\mu^aA_\mu^a\Biggl)\Biggr]}
{\int {\cal D}A_\mu^a\exp\Biggl[-\int dx\Biggl(\frac14
\left(\partial_\mu A_\nu^a-\partial_\nu A_\mu^a\right)^2+
\frac{g^2\eta^2}{3}A_\mu^{a2}\Biggr)\Biggr]}=$$

$$=\exp\Biggl\{-\frac{J^2q^2}{8\pi^2}\oint\limits_C^{}dx_\mu 
n^a(x)\oint\limits_C^{}dy_\nu n^a(y)\frac{1}{\left|x-y\right|}
\Biggl[\delta_{\mu\nu}\Biggl(\left(m+\frac{1}{m(x-y)^2}\right)
K_1+\frac{1}{2\left|x-y\right|}\left(K_0+K_2\right)\Biggr)-$$

$$-\frac{(x-y)_\mu (x-y)_\nu}{(x-y)^2}
\Biggl(3\Biggl(\frac{m}{4}
+\frac{1}{m(x-y)^2}\Biggr)K_1+\frac{3}{2\left|x-y\right|}
\left(K_0+K_2\right)+\frac{m}{4}K_3\Biggr)\Biggr]\Biggr\}, 
\eqno (4)$$
where $m\equiv\sqrt{\frac23}g\eta$ is the gauge bosons' mass, and 
$K_i$'s, $i=0,1,2,3,$ stand for the Macdonald functions, whose arguments 
are 
the same, $m\left|x-y\right|$. It is worth 
mentioning, that during this integration one should take 
into account that contrary to the Abelian case, where 

$$\partial_\mu j_\mu=0~ {\rm with}~ j_\mu(x)\equiv
\int\limits_0^1 d\tau\dot x_\mu(\tau)\delta(x-x(\tau)),$$ 
the current $j_\mu^a$ is not conserved, i.e. $\partial_\mu 
j_\mu^a\ne 0$.

Eq. (4) describes the interaction of the massive gauge bosons moving 
along the loop with each other, in the Gaussian approximation.    
The asymptotic behaviours of the R.H.S. of Eq. (4) at $m\sqrt{A}\ll 1$ and 
$m\sqrt{A}\gg 1$, where $A$ stands for the minimal area inside 
the contour $C$, whose square root is a typical distance 
between two points on $C$, read
 
$$\exp\Biggl\{-\frac{J^2q^2}{4\pi^2m^2}\oint\limits_C^{}dx_\mu 
\oint\limits_C^{}dy_\nu\frac{1}{\left|x-y\right|^4}
\Biggl[\delta_{\mu\nu}-\frac{4(x-y)_\mu 
(x-y)_\nu}{(x-y)^2}\Biggr]\Biggr\}$$
and

$$\exp\Biggl\{-\frac{J^2q^2\sqrt{m}}{8\sqrt{2}\pi^{\frac32}}\oint\limits_C^{}
dx_\mu n^a(x)\oint\limits_C^{}dy_\nu n^a(y)  
\frac{{\rm e}^{-m\left|x-y\right|}}{\left|
x-y\right|^{\frac32}}\Biggl[\delta_{\mu\nu}-\frac{(x-y)_\mu (x-y)_\nu}
{(x-y)^2}\Biggr]\Biggr\}$$ 
respectively.
 
The one-loop expression for the the Wilson loop average could be obtained 
upon accounting, in addition to the R.H.S. of Eq. (4), for the term 
resulting from the substitution of the saddle-point value of the 
Gaussian integral (4),   

$$A_\mu^{a{\,}{\rm extr.}}(p)=\frac{iJq}{p^2+m^2}\Biggl[j_\mu^a(p)+
\frac{p_\mu p_\nu}{m^2}j_\nu^a(p)\Biggr]$$
into the exponent 

$$\exp\Biggl[-g\varepsilon^{abc}\int dx\left(\partial_\mu A_\nu^a\right)
A_\mu^b A_\nu^c\Biggr].$$
However, this expression again obviously does not contain any 
nontrivial dependence on the string world-sheet. Therefore, at least
up to a one-loop approximation, in the Londons' limit, 
we are 
left only with the perimeter type contributions to the Wilson loop average. 

Let us now consider the Georgi-Glashow model 
without external charged particles. Then 
the nontrivial dependence on the world-sheet $S$ emerges 
in the weight factor $\left<{\cal F}(S)\right>$ of the string 
representation for the non-Abelian analog of the 't Hooft loop average,   
$\int {\cal D}x_\mu(\xi)\left<{\cal F}(S)\right>$. 
This weight factor has the form 

$$\left<{\cal F}(S)\right>=\frac{1}{{\cal Z}}
\int \left|\Phi^a\right| {\cal D}
\left|\Phi^a\right|{\cal D}A_\mu^a {\cal D}\theta^{{\rm reg.}}
\exp\Biggl\{-\int dx\Biggl[\frac14 F_{\mu\nu}^{a2}+\frac12\left|
D_\mu^{ab}\Phi^b\right|^2+\lambda\left(\left|\Phi^a\right|^2-
\eta^2\right)^2+$$

$$+\frac{\pi}{g}\varepsilon_{\mu\nu\alpha\beta}
T_{\mu\nu}F_{\alpha\beta}^a
N^{\prime a}\Biggr]\Biggr\}, \eqno (5)$$
where $\vec N^\prime$ is again a 
{\it constant} unit three-vector, independent 
of $\vec N$, which will drop out from the final expression 
for $\left<{\cal F}(S)\right>$. 
In what follows, our aim will be 
to find the correlators of the dual non-Abelian field strength 
tensors, which could be immediately done as soon as $\left<{\cal F}(S)
\right>$ is known, by applying to Eq. (5) the cumulant 
expansion$^{3}$. 

In the Londons' limit,  
after performing the duality transformation, Eq. (5) reads

$$\left<{\cal F}(S)
\right>=\frac{\int {\cal D}A_\mu^a\exp\Biggl\{-\int dx\Biggl[
\frac14 F_{\mu\nu}^{a2}+\frac{g^2\eta^2}{3}A_\mu^{a2}+\frac{\pi}{g}
\varepsilon_{\mu\nu\alpha\beta}T_{\mu\nu}F_{\alpha\beta}^a N^{\prime a}
\Biggr]\Biggr\}}{\int {\cal D}A_\mu^a\exp\Biggl\{-\int dx\Biggl[
\frac14 F_{\mu\nu}^{a2}+\frac{g^2\eta^2}{3}A_\mu^{a2}\Biggr]
\Biggr\}}. \eqno (6)$$
Within the one-loop approximation, Eq. (6) yields 

$$\left<{\cal F}(S)\right>=
\exp\Biggl\{\frac{1}{g^2}\int\limits_S^{}d\sigma_{\mu\nu}(x)
\int\limits_S^{} d\sigma_{\mu\nu}(y)\Biggl[\frac{\Lambda^4}{4}{\rm e}^{-
\frac{\Lambda^2(x-y)^2}{4}}-\frac{m^3K_1\left(m\left|x-y\right|\right)}
{\left|x-y\right|}\Biggr]-$$

$$-\frac{2m}{g^2}
\oint\limits_{\partial S}^{}dx_\mu\oint\limits_{\partial S}^{}dy_\mu
\frac{K_1\left(m\left|x-y\right|\right)}
{\left|x-y\right|}\Biggr\}, \eqno (7)$$
where the first term in the exponent on the R.H.S. of Eq. (7) is due 
to the regularization of the delta-function with $\Lambda$ 
standing for the  
UV cutoff, and the last term again vanishes, since $\partial S=0$. 
Comparing now Eq. (7) with the result of the cumulant expansion of Eq. (5) 
and parametrizing the 
bilocal correlator in the adjoint 
representation as follows$^{3}$

$$\left<\tilde F_{\mu\nu}^a(x)\tilde F_{\lambda\rho}^b(0)\right>=
\frac{\delta^{ab}}{3}\Biggl\{\Biggl(\delta_{\mu\lambda}\delta_{\nu\rho}-
\delta_{\mu\rho}\delta_{\nu\lambda}\Biggr)D\left(x^2\right)+$$

$$+\frac12
\Biggl[\partial_\mu\Biggl(x_\lambda\delta_{\nu\rho}-x_\rho\delta_{\nu\lambda}
\Biggr)+\partial_\nu\Biggl(x_\rho\delta_{\mu\lambda}-x_\lambda\delta_{\mu\rho}
\Biggr)\Biggr]D_1\left(x^2\right)\Biggr\},$$
we arrive at the following values of the functions $D$ and $D_1$

$$D\left(x^2\right)=\frac{3}{2\pi^2}\Biggl[\frac{m^3K_1\left(
m\left|x\right|\right)}
{\left|x\right|}-\frac{\Lambda^4}{4}{\rm e}^{-\frac{\Lambda^2x^2}{4}}
\Biggr], \eqno (8)$$

$$D_1\left(x^2\right)=\frac{3m}{\pi^2x^2}\Biggl[\frac{K_1\left(
m\left|x\right|\right)}{\left|x\right|}
+\frac{m}{2}\left(K_0\left(m\left|x\right|\right)
+K_2\left(m\left|x\right|\right)\right)\Biggr]. \eqno (9)$$
The asymptotic behaviours of these functions at $\left|x\right|\ll
\frac{1}{m}$ and $\left|x\right|\gg\frac{1}{m}$ read

$$D\longrightarrow\frac{3m^2}{2\pi^2x^2}, \eqno (10)$$

$$D_1\longrightarrow\frac{6}{\pi^2\left|x\right|^4}, \eqno (11)$$
and

$$D\longrightarrow\frac{3m^4}{2\sqrt{2}\pi^{\frac32}}\frac{{\rm e}^{-m
\left|x\right|}}{\left(m\left|x\right|\right)^{\frac32}}, \eqno (12)$$

$$D_1\longrightarrow\frac{3m^4}{\sqrt{2}\pi^{\frac32}}\frac{
{\rm e}^{-m\left|x\right|}}{\left(m\left|x\right|\right)^{\frac52}}
\eqno (13)$$ 
respectively, where during the derivation of Eq. (10) we have put 
$\Lambda$ to be of the order of $m$, which is true in the confining 
regime$^{1}$. 

Thus, we conclude that the bilocal approximation to MVC is exact in the 
Londons' limit of the Georgi-Glashow model in the one-loop approximation, 
i.e. analogously to the AHM in the 
Londons' limit$^{1}$, all the cumulants higher 
than the quadratic one vanish. 
Coefficient functions (8) and (9) are the main result of the present 
Letter. Similar to the Abelian case$^{1}$, 
their asymptotic behaviours (10)-(13) are in a good agreement
with the observable lattice data$^{4}$ on the corresponding behaviours 
in gluodynamics.

\vspace{6mm}
{\large \bf Acknowledgments} 
\vspace{3mm}

The author is grateful to Professors M.I.Polikarpov and Yu.A.Simonov 
for the encouragement and to  
M.N.Chernodub for criticism and a lot of useful discussions 
on the string representation of AHM. He would also like to thank the 
Quantum Field Theory Department of the Institute of Physics of the 
Humboldt University of Berlin and especially Profs. D.Ebert, D.L\"ust, 
and M.M\"uller-Preussker for kind hospitality.

\vspace{6mm}
{\large \bf References}

\vspace{3mm}
\noindent
1.~D.V.Antonov, {\it hep-th}/9710144.\\
2.~D.V.Antonov, {\it hep-th}/9710194.\\
3.~H.G.Dosch, {\it Phys.Lett.} {\bf 190}, 177 (1987); Yu.A.Simonov, 
{\it Nucl.Phys.} 
{\bf B307}, 512 (1988); H.G.Dosch and Yu.A.Simonov, {\it Phys.Lett.} 
{\bf B205}, 
339 (1988), {\it Z.Phys.} {\bf C45}, 147 (1989); Yu.A.Simonov, 
{\it Nucl.Phys.}  
{\bf B324}, 67 (1989), {\it Phys.Lett.} {\bf B226}, 151 (1989), 
{\it Phys.Lett.}  
{\bf B228}, 413 (1989), {\it Yad.Fiz.} {\bf 58}, 113 (1995), preprint 
ITEP-PH-97-4 
({\it hep-ph}/9704301), {\it hep-ph}/9712248, {\it hep-ph}/9712250; 
for a review see  
Yu.A.Simonov, {\it Yad.Fiz.} {\bf 54}, 192 (1991), {\it Phys.Usp.}  
{\bf 39}, 313 (1996).\\
4.~A. Di Giacomo and H.Panagopoulos, {\it Phys.Lett.} {\bf B285}, 
133 (1992); 
A. Di Giacomo, E.Meggiolaro, and H.Panagopoulos, {\it hep-lat}/9603017 
(preprints IFUP-TH 12/96 and UCY-PHY-96/5) (in press in {\it Nucl.Phys.} 
{\bf B}).\\
5.~K.Lee, {\it Phys.Rev.} {\bf D48}, 2493 (1993).\\
6.~M.I.Polikarpov, U.-J.Wiese, and M.A.Zubkov, {\it Phys.Lett.} 
{\bf B309}, 133 (1993); P.Orland, {\it Nucl.Phys.} {\bf B428}, 221 (1994);  
M.Sato and S.Yahikozawa, {\it Nucl.Phys.} {\bf B436}, 100 (1995); 
E.T.Akhmedov, M.N.Chernodub, M.I.Polikarpov, and M.A.Zubkov, 
{\it Phys.Rev.} {\bf D53}, 2087 (1996).\\ 
7.~E.T.Akhmedov, {\it JETP Lett.} {\bf 64}, 82 (1996).\\
8.~D.I.Diakonov and V.Yu.Petrov, {\it Phys.Lett.} {\bf B224}, 131 
(1989).\\
9.~D.V.Antonov and D.Ebert, {\it Mod.Phys.Lett.} {\bf A12}, 2047 
(1997).\\ 
10.D.V.Antonov, {\it hep-th}/9707245.\\

\end{document}